\documentclass[prl,twocolumn,showpacs,preprintnumbers,amsmath,amssymb]{revtex4}
\usepackage{graphicx}
\usepackage{dcolumn}
\usepackage{bm}
\usepackage{psfig}
\newcommand \be{\begin{eqnarray}}
\newcommand \ee{\end{eqnarray}}
\newcommand {\ket}[1]{|#1\rangle}
\newcommand {\bra}[1]{\langle #1|}
\begin{document}
           \csname @twocolumnfalse\endcsname
\title{Enhancement of pairing due to the presence of resonant cavities}
\author{K. Morawetz$^{1,2}$, B. Schmidt$^1$, M. Schreiber$^1$, P. Lipavsk\'y$^{3,4}$}
\affiliation{$^1$Institute of Physics, Chemnitz University of Technology, 
09107 Chemnitz, Germany}
\affiliation{$^2$Max-Planck-Institute for the Physics of Complex
Systems, Noethnitzer Str. 38, 01187 Dresden, Germany}
\affiliation{$^3$Faculty of Mathematics and Physics, Charles University, Ke Karlovu 5, 12116 Prague 2}
\affiliation{
$^4$Institute of Physics, Academy of Sciences, 
Cukrovarnick\'a 10, 16253 Prague 6, Czech Republic}

\begin{abstract}
A correlated fermion system is considered surrounding a finite cavity with virtual levels. The pairing properties are calculated and the influence of the cavity is demonstrated. To this end the Gell-Mann and Goldberger formula is generalized to many-body systems. We find a possible enhancement of pairing temperature if the Fermi momentum times the cavity radius fulfills a certain resonance condition which suggests an experimental realization. 
\end{abstract}
\maketitle

Correlation effects in finite nanostructures are paid growing attention since it is hoped to exploit perhaps new transport phenomena. So far most treatments concentrate on one-particle properties like electron current and heat conduction through such devices. The sophistication of experimental devices has reached such a level that one can anticipate also to exploit two-particle correlation effects in finite quantum structures. The most prominent one in correlated many-body systems is certainly the occurrence of pairing and superconductivity which  one can now study on quantum dots. 

There is fast experimental progress on studying pairing properties in such finite systems. If a finite system with discrete levels is coupled to a continuum, a particle scattered on the system can be trapped for a certain time in such level forming Feshbach resonances. The condensation of pairs near a Feshbach resonance has been observed \cite{RGJ04,ZSSRKK04} analogously to superconductivity. The fermion condensation using ultracold atomic $^6$Li or $^{40}$K clouds confirmed that the BCS superfluid state has been reached \cite{JBAHRCDG03,GRJ03}. There is also clear experimental evidence for superfluidity in a resonantly interacting Fermi gas \cite{KHGTT04} which arises near a Feshbach resonance.  Though two-body physics does not allow the occurrence of bound states there, the many-body effects allow such pairing for fermionic atoms. Consequently there is a growing theoretical activity \cite{CKBZKTTKS04} with suggestions even that the Fulde-Ferrell-Larkin-Ovchinnikov state might be observable due to spatially modulated superfluid phases in atomic fermion systems \cite{MMI04}.   

Though both fields, the transport in nanostructures and the two-particle correlations like superconductivity in many-body systems, are heavily explored separately, the combined problem is rarely attacked \cite{CKBZKTTKS04}. The most elaborate treatment has been performed within a related problem of crystal fields \cite{F79} and mechanisms of pair breaking \cite{FZ89}. In this paper we want to explore the pairing properties of the many-body system in the presence of a finite nanostructure and will show how the pairing and critical temperature are changed.

One can consider such finite quantum structure as a cavity fixed relative to the much lighter electrons surrounding this cavity. We will solve this problem of two-electron pairing in their medium together with the cavity not in the Faddeev language \cite{Be72} but by coupled-channel scattering theory \cite{T72} in the Gell-Mann and Goldberger formulation \cite{GG53}. This will result in a formula of the total ${\cal T}$-matrix of two interacting particles with external interaction in terms of two separated problems: (i) two particles interacting only with themselves and (ii) two particles interacting only with an external potential. To the author's knowledge this treatment has not been extended from two-particle scattering to many-particle correlations in a surrounding medium. In the first part we extend therefore the Gell-Mann and Goldberger formula to many-body systems with the help of Green's functions analogously to the derivation without medium \cite{MSSFL04}. This formula is then solved in the second part of the paper for specific model interactions between the two electrons and between the electrons and the cavity.

We will consider the many-body system as well described by quasiparticles with energy $\varepsilon_p$ and a momentum-dependent distribution function $f_p$. The cavity is assumed not to influence the overall homogeneous distribution of the medium. Denoting the center-of-mass momentum of two particles with capital letters, $P=p_1+p_2$ and the difference momentum with small letters $p=(p_1-p_2)/2$, the free retarded two-particle Green's function describing two freely moving quasiparticle in a medium reads
\be
\bra{pP} {\cal G}_0(\omega,t)\ket{p'P'}&=&(2\pi\hbar)^6 \delta(P-P')\delta(p-p')\nonumber\\&&\times {1-f_{\frac P 2 +p}-f_{\frac P 2 -p}\over \hbar \omega-\varepsilon_{\frac P 2 +p}-\varepsilon_{\frac P 2 -p} +i \eta}
\label{g0}
\ee
with an infinitesimal $\eta$ ensuring the causality and retarded character of the function. The Fourier transform $\hbar \omega$ of the difference time between the beginning and the end of the propagation describes the energy of possible excitations in this system. The quasiparticle distribution functions $f_p$ and energies $\varepsilon_p$ in (\ref{g0}) represent the effect of the correlated medium surrounding the two particles.

Now we consider the interactions $V_{01}$ and $V_{02}$ of the two particles with the external cavity. The correlated two-particle Green's function in ladder approximation is then given by the integral equation
\be
{\cal G}_1={\cal G}_0+{\cal G}_0 (V_{01}+V_{02}) {\cal G}_1
\label{g}
\ee
where we used operator notation understanding products as integration about intermediate variables. The interaction of both particles with the cavity is collected in a first channel $V_1=V_{01}+V_{02}$ and the interaction between the two particles is described by a second channel $V_2=V_{12}$. Then we can define channel Green's functions  \be
{\cal G}_i(\omega) = {\cal G}_0(\omega)+{\cal G}_i(\omega) V_i {\cal G}_0(\omega)
\label{calG}
\ee
corresponding to the $i$th-channel ${\cal T}$-matrix
${\cal T}_i(\omega)=V_i+V_i {\cal G}_i(\omega) V_i.
$
The total ${\cal T}$-matrix can be written as a sum of auxiliary ${\cal T}$-matrices 
\be
{\cal T}(\omega)=\sum\limits_i V_i + \sum\limits_i V_i {\cal G}_0(\omega) {\cal T}(\omega) = \sum\limits_i {\cal T}'_i(\omega)
\label{sumT}
\ee
which read
\be
{\cal T}'_i&=&V_i+V_i {\cal G}_0 {\cal T}
=
{\cal T}_i(\omega) + \sum\limits_{j \neq i} {\cal T}_i(\omega) {\cal G}_0(\omega) {\cal T}'_j(\omega)
\label{Ts}
\ee
where we used (\ref{calG}) and $V_i {\cal G}_i = {\cal T}_i {\cal G}_0$ to derive the equality. Intro\-du\-cing ${\cal T}_1'$ into ${\cal T}_2'$  in (\ref{Ts}) leads to
${\cal T}_2'={\cal T}_2(1+{\cal G}_0{\cal T}_1)+{\cal T}_2 {\cal G}_0 {\cal T}_1 {\cal G}_0 {\cal T}_2'
$
with the help of which we define
\be
{\cal T}_{ab}\equiv {\cal T}_2' (1+{\cal G}_0 {\cal T}_1)^{-1}&=&{\cal T}_2+{\cal T}_2 {\cal G}_0 {\cal T}_1 {\cal G}_0 {\cal T}_{ab}.
\label{ab}
\ee
Using again (\ref{calG}) one obtains
\be
(1-V_2{\cal G}_1)^{-1}V_2&=&(1-{\cal T}_2{\cal G}_0V_1{\cal G}_1)^{-1} {\cal T}_2
\label{zv}
\ee
and the right hand side of (\ref{zv}) is just the definition of ${\cal T}_{ab}$ from (\ref{ab}) using once more ${\cal T}_1{\cal G}_0=V_1 {\cal G}_1$. Therefore we see that from (\ref{zv}) the equation for ${\cal T}_{ab}$ follows,
\be
{\cal T}_{ab}=V_2+V_2 {\cal G}_1 {\cal T}_{ab}.
\label{tab}
\ee
The total ${\cal T}$-matrix (\ref{sumT}) can then be written with (\ref{Ts}) as
\be
{\cal T}&=&{\cal T}_1+(1+{\cal T}_1{\cal G}_0) {\cal T}_2'
\nonumber\\
&=&{\cal T}_1+(1+{\cal T}_1{\cal G}_0) {\cal T}_{ab} (1+{\cal G}_0 {\cal T}_1).
\label{GMa}
\ee
This formula together with (\ref{tab}) and (\ref{g}) is the Gell-Mann and Goldberger formulation now generalized to correlated many-body systems via (\ref{g0}).  The original Gell-Mann and Goldberger formula is exact for the scattering of two particles in the presence of a third potential and is equivalent to the Faddeev equation. Our generalization to many-body systems treat the correlation effects on the level of the ladder approximation.

The strategy is to determine first the correlated two-particle Green's function in the presence of the cavity (\ref{g}) in order to obtain the correlated two-particle ${\cal T}$-matrix (\ref{tab}) in the two-particle channel. The total ${\cal T}$-matrix is then constructed from (\ref{GMa}).

Since we are interested in the pairing properties we will search for the onset of pairing as a critical temperature where the correlated ${\cal T}$-matrix (\ref{tab}) has poles at twice the chemical potential according to the Thouless criterion. Therefore it is sufficient for our purpose to solve (\ref{tab}). The necessary correlated two-particle Green's function (\ref{g}) can be given provided we know the solution of the single-particle problem of an electron and the cavity, $(p_1^2/2m+V_{01})\ket{n_1}=E_{n_1}\ket{n_1}$. Assuming this to be the case, the related two-particle problem separates
\be
({p_1^2\over 2m_1}\!+\!{p_2^2\over 2m_2}\!+\!V_{01}\!+\!V_{02})\ket{n_1n_2}\!=\!(E_{n_1}\!+\!E_{n_2})\ket{n_1n_2}.
\label{schro}
\ee
We can project (\ref{g}) onto the complete set of wave functions $\ket{n_1 n_2}$ and obtain from (\ref{g})
the solution
\be
\bra{pP}{\cal G}_1\ket{p'P'}\!&=&\!\sum\limits_{n_1n_2}{\bra{pP} n_1 n_2 \rangle (1\!-\!f_{n_1}\!-\!f_{n_2}) \bra{n_1n_2} p'P'\rangle 
\over \hbar \omega-E_{n_1}-E_{n_2}+\Delta E_n+i\eta}
\nonumber\\&&
\label{g1}
\ee
with
\be
\Delta E_n &=& {p_{n_1}^2\over 2 m_1}\!+\!{p_{n_2}^2\over 2 m_2}\!-\!\varepsilon_{n_1}\!-\!\varepsilon_{n_2}
\nonumber\\&+&(E_{n_1}\!+\!E_{n_2}\!-\!{p_{n_1}^2\over 2 m_1}\!-\!{p_{n_2}^2\over 2 m_2}\!)(f_{n_1}\!+\!f_{n_2}).
\label{de}
\ee

We end up with the same formula as in \cite{MSSFL04} but with in-medium effects represented by the distribution functions $f$ and the quasiparticle energy $\varepsilon$
\be
&&\bra{pP}{\cal T}_{ab}\ket{p'P'}=V_{pp'}\delta_{PP'}+\sum\limits_{n_1n_2 \bar p \bar p'\bar{P}}V_{p\bar p}(1\!-\!f_{n_1}\!-\!f_{n_2})
\nonumber\\
&&\times {\langle \bar pP |n_1n_2\rangle \langle n_1n_2| \bar p'\bar{P}\rangle \over \hbar \omega-E_{n_1}-E_{n_2}+\Delta E_n+i\eta} \bra{\bar p'\bar{P}}{\cal T}_{ab}\ket{p' P'}.
\label{tmat}
\ee
We will solve the ${\cal T}$-matrix (\ref{tmat}) with the help of a separable potential
$V_{pp'}=\lambda g_p g_p'$ between the two particles \cite{Y59}. Any finite range potential can be represented by a finite-rank separable potential \cite{Ko92,KPML97}. Here we restrict ourselves to a rank-one potential $g_p=1/(p^2+\beta^2)$ which is sufficient to describe the s-wave pairing interaction near the Fermi surface. 

As a model for the cavity we will choose an opaque wall  ${V}_{01}(r)={\hbar^2\over 2 m} {\Omega \over R} \delta (r-R)$ with a coupling strength $\Omega$ and a finite radius $R$. Then the radial single-particle wave function \cite{F94}
\be
\chi_k(r)=A_{x} \sin{k r} \left \{ \begin{array}{ll} 1 & r<R \cr 1+{\Omega \sin{x}\over x}  {\cos{(x+k r)}\over \sin{k r}}& r>R\end{array}\right \}
\ee
with momenta $p=\hbar k$ and $x=k R$ differs from the plane wave, near the cavity or alternatively for $\Omega\ll k R$ only by an amplitude \cite{F94}
\be
A_x^2={1+\tan^2{x}\over \tan^2{x}+(1+{\Omega \over x}\tan{x})^2}.
\label{A1}
\ee
Since the two-particle Schr{\"o}dinger equation (\ref{schro}) with $V_{01}+V_{02}$ separates
we have approximately
$\ket{n_1n_2}\approx A_{|K/2-k|R}A_{|K/2+k|R}\ket{pP}
$
with plane waves $\ket{pP}$ and the center-of-mass momentum $P=\hbar K=p_1+p_2$ and relative momentum $p=\hbar k=p_1-p_2$. It is worth noting that the normalization of plane waves to the current remains unchanged by the amplitude $A_x$ since from (\ref{A1}) we have $\lim\limits_{x\to\infty} A_x^2=1$. The solution of (\ref{tmat}) then reads
\be
\bra{p_1P}{\cal T}_{ab}\ket{p_2 P'}=\delta_{PP'} {\lambda g_{p_1}g_{p_2}\over 1-\lambda J(P,\omega)} 
\label{sol}
\ee
with 
\be
J(P,\omega)&=&\sum\limits_p g_p^2 
{|A_{|{K\over 2}+k|R}A_{|{K \over 2}-k|R}|^2 (1\!-\!f_{{P\over 2}\!+\! p}\!-\!f_{{P\over 2}\!-\!p})\over \omega -E_{{P\over 2}+p}-E_{{P\over 2}-p}+\Delta E+i\eta}.
\label{J}
\nonumber\\&&
\ee
For our opaque-wall cavity $E_{n}={p_{n}^2\over 2 m}$ holds and  (\ref{de}) simplifies to $\Delta E={({\frac{P}{ 2} \!+\!p})^2\over 2 m_1}+{({\frac{P}{ 2} \!-\!p})^2\over 2 m_2}\!-\!\varepsilon_{\frac{P}{ 2} \!+\!p}\!-\!\varepsilon_{\frac {P}{ 2} \!-\!p}$ which vanishes if we neglect renormalizations due to quasiparticle energies.

Before discussing this result we want to consider
two limiting cases.
Neglecting the wave-function renormalization $A\to1$ we obtain the standard expression for two-particle scattering in an infinite extended medium with separable interaction \cite{SRS90,ARS94,MOR94,KK97}. With (\ref{sol}) and (\ref{J}) we present the ${\cal T}$-matrix of the two particles in the presence of the cavity and in the medium which generalizes approaches without the cavity.
Oppositely, if we neglect the medium effects $f\approx0$ we obtain an approximation for the two-particle cavity with a separable interaction \cite{MSSFL04}. Expression (\ref{J}) without medium effects represents a much more convenient form than the numerically more elaborate equation (57) of \cite{MSSFL04} and it agrees with the latter up to some percent.

\begin{figure}
{\psfig{file=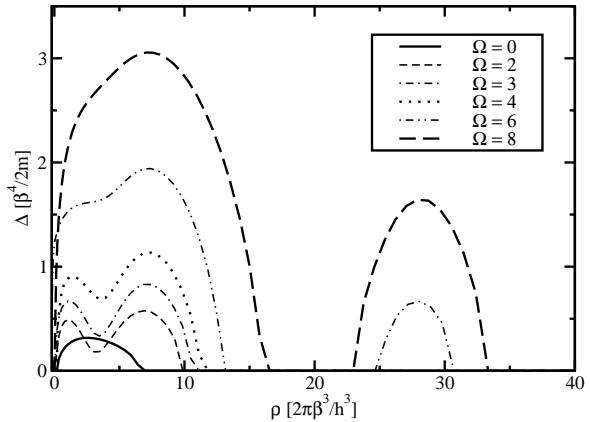,width=9cm}}
\vspace*{-3ex}
\caption{The gap $\Delta$ versus density for different cavity strengths $\Omega$, temperature $T=0.05 \beta^2/2m$, and form factors $g_p=1/(p^2+\beta^2)$. The parameters are chosen such that the scattering length is $a_0=-21.5 \hbar/\beta$, the free binding energy $E_b^0=0.005 \beta^2/2m$, and the radius of the cavity $R=4\hbar/\beta$.  
}\label{fig1}
\end{figure}


Now we investigate the pairing poles of the ${\cal T}$-matrix at energies twice the chemical potential and zero center-of-mass momentum. The ${\cal T}$-matrix separates in momenta near this pole and we find the gap equation 
\be
\Delta(p)=-\lambda \sum\limits_{\bar p} g_pg_{\bar p} A_{p R}^2 A_{\bar p R}^2 {\tanh{E_{\bar p}} \over 2 E_{\bar p}} \Delta(\bar p)
\label{gap}
\ee
with $E_p=\sqrt{(\varepsilon_p-\mu)^2+\Delta(p)^2}$. The momentum dependence of the gap follows as $\Delta(p)=g_p A_{pR}^2 \Delta$ with $\Delta$ found from (\ref{gap}) plotted in figure~\ref{fig1}. We see that the influence of the resonant cavity is enhancing the gap and inducing maxima in the density dependence. It is remarkable that a second branch of superconductivity appears at higher densities due to the presence of the resonant cavity.

The additional factors in (\ref{gap}) compared to the free gap equation are just the correlation function within the theory of pair breaking \cite{dG89} but here extended by the potential form factors $g_p$.

\begin{figure}
{\psfig{file=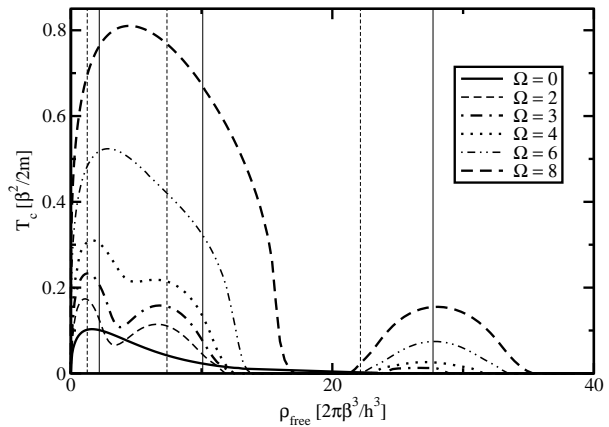,width=9cm}}
\vspace*{-3ex}\caption{The critical temperature $T_c$ 
versus density for the same parameters as in figure~\protect\ref{fig1} except that the temperature is at $T=T_c$. The perpendicular lines indicate the minima of $A_{k_f R}^4$ for $k_{f} R \gg \Omega$ (solid) and for $k_f R \ll \Omega$ (dashed) according to (\protect\ref{cond}).
}\label{fig2}
\end{figure}

The critical temperature following from (\ref{gap}) for $\Delta = 0$ is plotted in figure~\ref{fig2} where one sees that it is enhanced due to the presence of the cavity compared to the case without cavity $\Omega=0$. Interestingly again there appears a second separated branch for higher densities where pairing occurs. 

From the pole of the ${\cal T}$-matrix (\ref{sol}) at negative frequencies we can deduce the binding energy of two particles. With increasing density the bound state energy is decreasing until it vanishes at the Mott density. This behavior is strongly influenced by the cavity strength.
We have chosen a situation where the cavity is enhancing pairing as well as binding. This property depends on the value of the scattering length $a_0$ and the cavity radius $R$; we refer to \cite{MSSFL04} for details.  

The appearance of maxima in the critical temperature can be approximately described as a modification of the usual BCS theory due to the cavity. For this purpose we consider the potential strength at the Fermi surface $V_0=\lambda g_{p_f}^2$ which can be taken out of the sum in (\ref{gap}) introducing at the same time an energy cutoff $\omega_c$, which is not necessary when taking into account the complete finite range $g_p$ as done above. Then the standard procedure to extract the critical temperature in BCS theory applies and using the free density of states $N$ at the Fermi energy we obtain
\be
T_c=1.13 \omega_c {\rm e}^{-{2\over N V_0 A_{k_f R}^4}}=\left ({T_{c}^0\over 1.13 \omega_c}\right )^{1\over A_{k_f R}^4}.
\label{tc}
\ee
This means that the critical BCS temperature without cavity $T_{c}^0$ is modified by the amplitude $A_{k_f R}$ at the Fermi momentum. This result is in agreement with the Anderson theorem
\cite{Ric69a} which states, that for a homogeneous
perturbation and order parameter the critical temperature can only be
effected by the density of states. Here the amplitude of the cavity effects the density of states. The critical temperature can be enhanced if $x=k_f R$ takes values where the cavity amplitude (\ref{A1}) has minima. These minima occur at $2 x_n=n \pi -\arctan{2x_n/\Omega}$ with the values \cite{F94}
\be
\left ({1\over A_{x}^4}\right )_{max}\approx 
\left \{ 
\begin {array}{ll}
\left ({2 \Omega \over n \pi}\right )^4 & x_n=n {\pi \over 2}< \Omega \cr
1+{8 \Omega \over (2 n-1 )\pi} & x_n=(2 n-1) {\pi \over 4}> \Omega
\end{array}
\right .
\label{cond}
\ee
and $n=1,3..$. In figure~\ref{fig2} these points are indicated by perpendicular lines and agree nearly with the observed maxima.

The parameter choice of such resonant cavities seems to be a reasonable way to enhance the pairing temperature as the following simple estimate shows. 
Assuming the Fermi energy to be about $1$eV and the Fermi momentum just fulfilling the first resonance condition $k_f R=\pi \hbar /2$, the required cavity radius would be about $200$nm which is realistic to be fabricated.  The enhancement of the BCS critical temperature (\ref{tc}) could then be remarkable as seen in figure~\ref{fig2}.

Summarizing we have investigated the modification of the pairing temperature and the range of superconductivity due to the presence of a resonant cavity. We suggest to construct an experiment with interacting electrons possessing an effective attracting interaction due to background phonon coupling and an additional resonant cavity. If the cavity radius $R$ and the cavity strength $\Omega$ are chosen such that the condition ${k_f R}=n\pi/2<\Omega$ or ${k_f R}=(2 n-1) \pi/4 >\Omega$ for $n=1,3,5..$ are fulfilled we expect a remarkable enhancement of the critical temperature. We expect the effect described here for two-particle pairing properties to remain also for a macroscopic number of coherent paired particles since the coherence of such state can be kept if many of such resonant cavities are arranged within a regular crystal structure.

The helpful discussions with P. Fulde are gratefully acknowledged.
\bibliography{kmsr,kmsr1,kmsr2,kmsr3,kmsr4,kmsr5,kmsr6,kmsr7,delay2,spin,refer,delay3,gdr,chaos,sem3,sem1,sem2,spin1}

\end{document}